\documentclass[conference]{IEEEtran}
\usepackage{cite}
\usepackage{graphicx}
\usepackage{amsmath}
\usepackage{array}
\usepackage{url}
\usepackage{booktabs}
\usepackage[hidelinks]{hyperref}  
\usepackage{balance}              

\usepackage{xspace}

\title{Dual‐Path Phishing Detection: \\Integrating Transformer‐Based NLP with \\Structural URL Analysis}

\author{
\IEEEauthorblockN{Ibrahim Altan, Abdulla Bachir, Yousuf Parbhulkar, Abdul Muksith Rizvi, Moshiur Farazi}
\IEEEauthorblockA{College of Computing and IT, University of Doha for Science and Technology, Doha, Qatar \\
ibrahimaltan243@gmail.com, Abdulla\_170b@outlook.com, yousef.zahid@gmail.com, \\
abdulmuksith3@yahoo.com, moshiur.farazi@udst.edu.qa}

}

\begin{document}

\maketitle

\begin{abstract}
Phishing emails pose a persistent and increasingly sophisticated threat, undermining email security through deceptive tactics designed to exploit both semantic and structural vulnerabilities. Traditional detection methods, often based on isolated analysis of email content or embedded URLs, fail to comprehensively address these evolving attacks. In this paper, we propose a dual-path phishing detection framework that integrates transformer-based natural language processing (NLP) with classical machine learning to jointly analyze email text and embedded URLs. Our approach leverages the complementary strengths of semantic analysis using fine-tuned transformer architectures (e.g., DistilBERT) and structural link analysis via character-level TF-IDF vectorization paired with classical classifiers (e.g., Random Forest). Empirical evaluation on representative email and URL datasets demonstrates that this combined approach significantly improves detection accuracy. Specifically, the DistilBERT model achieves a near-optimal balance between accuracy and computational efficiency for textual phishing detection, while Random Forest notably outperforms other classical classifiers in identifying malicious URLs. The modular design allows flexibility for standalone deployment or ensemble integration, facilitating real-world adoption. Collectively, our results highlight the efficacy and practical value of this dual-path approach, establishing a scalable, accurate, and interpretable solution capable of enhancing email security against contemporary phishing threats.
\end{abstract} 

\begin{IEEEkeywords}
Phishing Detection, Natural Language Processing (NLP), Transformer Models, URL Analysis, Machine Learning, Email Security
\end{IEEEkeywords}

\section{Introduction}
Email continues to serve as a critical communication medium for individuals and businesses worldwide. Due to its widespread adoption has made email a prime target for various cyber threats, particularly phishing attacks. According to recent statistics, approximately 3.4 billion phishing emails are sent daily worldwide highlighting the severity and scale of this threat \cite{aag_it_phishing_stats}. Recent advancements in Natural Language Processing (NLP) and Large Language Models (LLMs) have further enabled cybercriminals to craft highly sophisticated and contextually accurate phishing emails \cite{kyaw2024}. Traditional phishing detection methods, such as signature-based and rule-based approaches, often fail to effectively combat modern email threats because of their static nature and limited ability to generalize to novel and evolving attacks \cite{thakur2023}. Additionally, malicious actors themselves now increasingly leverage advanced NLP and machine learning techniques to evade traditional security measures and bypass detection systems \cite{atlam2023}. This landscape has heightened the need for more sophisticated, intelligent detection solutions.


\begin{figure}[t] 
    \centering
    \includegraphics[width=\linewidth]{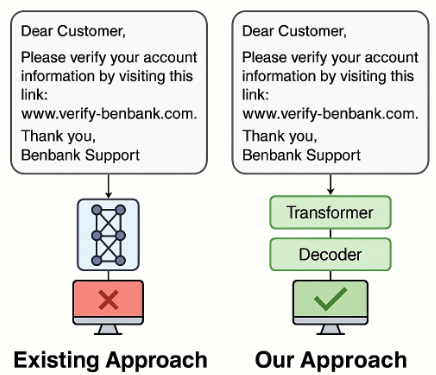}
    \caption{Illustrative comparison of a conventional classifier and our Transformer‐based detector on the same spoofed email (“www.verify‐benbank.com”). The baseline model (left) fails to recognize the fraudulent link and misclassifies the message, whereas our encoder–decoder Transformer (right) correctly flags it as phishing.}

    \label{fig:teaser}
\end{figure}

To address the growing complexity of phishing emails, this paper leverages transformer-based models such as BERT and DistilBERT, which are well-suited for understanding the semantic and contextual nuances of email content. 
They can be fine-tuned with relatively few labeled examples, making them highly effective for adapting to novel phishing variants and zero-day attacks. 

In response to these challenges, our paper aims to develop and implement an advanced AI-based phishing email detection model. Specifically, we propose integrating NLP and deep learning techniques to analyze and classify email content effectively and evaluate embedded links for malicious intent. Our goal is not only to enhance the accuracy and effectiveness of phishing detection but also to explore the feasibility and performance benefits of combining both content-based classification and link analysis into a unified detection framework \cite{rayala2023, kyaw2024}.


The remainder of this paper is structured as follows: Section II provides background information and foundational concepts; Section III reviews existing related work and identifies current research gaps; Section IV outlines our proposed integrated solution and methodology; Section V presents the experimental evaluation and results; Section VI discusses the insights derived from our AI-based model; and Section VII concludes by summarizing the project's contributions and future research directions.

\section{Related Works}

Early spam and phishing detection relied heavily on classical machine learning and statistical approaches. One of the foundational works in this area by Sahami et al. introduced a Naïve Bayes classifier for email filtering, using word frequency features to distinguish between spam and ham \cite{sahami1998}. Similarly, Drucker et al. applied Support Vector Machines (SVMs) to large-scale spam datasets and demonstrated superior accuracy over rule-based systems \cite{drucker1999}. These early approaches laid the groundwork for automated email classification by proving the viability of machine learning on textual features. However, they relied heavily on handcrafted features and were limited in their ability to generalize to more sophisticated phishing tactics.

The introduction of contextual word embeddings and transformer architectures significantly improved phishing detection. BERT (Bidirectional Encoder Representations from Transformers) emerged as a highly effective model due to its ability to understand word dependencies and context \cite{zhang2020}. Researchers like AbdulNabi and Yaseen showed that fine-tuning BERT embeddings outperformed conventional classifiers in detecting spam and phishing emails \cite{abdulnabi2021}. These models not only improved accuracy but also offered adaptability to evolving attack vectors with minimal retraining.

Despite these advancements, phishing attacks increasingly employ deceptive tactics where the email content appears legitimate, but embedded URLs redirect users to malicious sites. To address this, hybrid detection frameworks combining email body analysis with URL inspection have been proposed. Sahingoz et al. integrated NLP-based content features with lexical URL features (e.g., domain length, presence of IP addresses), demonstrating that combining both signals significantly enhances classification robustness \cite{sahingoz2019}. This hybrid approach is particularly effective against emails that bypass text-based models by embedding weaponized links in otherwise benign messages.

Our work builds upon this hybrid model concept by using fine-tuned transformer models for content analysis and TF-IDF vectorization with Random Forest and Logistic Regression for URL classification. In doing so, it bridges the gap between NLP-driven semantic understanding and structural URL analysis. Moreover, our integration of explainability tools such as attention weight visualization and LIME contributes to increased transparency—an often overlooked but critical component in real-world phishing defense systems.

\subsection{Natural Language Processing}

Natural Language Processing (NLP) has emerged as a powerful technique in phishing detection, particularly in analyzing the linguistic characteristics of phishing emails. Unlike traditional rule-based filters, NLP-based systems can detect more subtle cues hidden in text, such as grammatical inconsistencies, stylometric anomalies, and context deviations.

\subsubsection{Structural and Stylometric Feature-Based Detection}

One of the earliest and most influential studies in this domain was conducted by Chandrasekaran et al. \cite{chandrasekaran2006}, who proposed a phishing email detection framework based on stylometric and structural features. Their approach was grounded in the observation that phishing emails often exhibit significantly different linguistic properties compared to legitimate emails. The researchers extracted features such as sentence length, punctuation frequency, function word usage, and lexical richness. These were then input into a supervised decision tree classifier.

Their results indicated that phishing emails typically contain shorter sentences and reduced vocabulary diversity, suggesting automated or careless generation. Although their method did not rely on semantic-level modeling, it laid foundational groundwork for future NLP-based phishing detection techniques by demonstrating that even surface-level textual analysis can achieve high precision and recall in phishing classification tasks.

\subsubsection{Lexical and Semantic Enrichment with Machine Learning}

Building on these early findings, Sahingoz et al. \cite{sahingoz2019} presented an advanced framework that combined lexical, syntactic, and semantic features to enhance phishing detection. Their model incorporated a robust preprocessing pipeline involving tokenization, lemmatization, stemming, and stop-word removal, followed by the use of TF-IDF vectors and n-gram models (including uni-, bi-, and tri-grams) to represent email text numerically.

To evaluate performance, they tested multiple classifiers, including Support Vector Machines (SVM), k-Nearest Neighbors (k-NN), Naïve Bayes, and Random Forest. The Random Forest model exhibited the highest accuracy and resilience across various phishing strategies. 

\subsubsection{Deep Contextual Modeling using Transformers}

Recent advancements in deep learning and transformer models have further revolutionized phishing detection. Zhang et al. \cite{zhang2020} introduced a BERT-based framework capable of understanding deep contextual relationships in email text. 

In their approach, emails were preprocessed using standard NLP techniques, after which the text was passed through a pre-trained BERT model. The [CLS] token representation was extracted from BERT's final hidden layer and fed into a dense neural layer with SoftMax activation for binary classification. Their experiments demonstrated that the BERT-based model outperformed both traditional classifiers and shallow neural networks in precision, recall, and F1-score.


\subsubsection{Integration of URL Analysis and NLP Features}

Another critical innovation in this space is the combined use of NLP and URL analysis. Modern phishing emails often embed malicious links that are difficult to detect using body text alone. By incorporating features like domain length, use of IP addresses, and presence of obfuscating characters (e.g., “0” for “O”), these hybrid systems can detect threats that traditional spam filters or content-only NLP models might overlook.

Sahingoz et al.'s framework notably integrated URL-based features with text analysis, highlighting that phishing emails often rely on both linguistic deception and link-based redirection. This comprehensive strategy significantly enhanced model performance and reduced the risk of false negatives in sophisticated phishing attempts.

\subsection{Malicious URL Detection Methods}

While content-based analysis plays a key role in phishing detection, many modern attacks rely heavily on malicious URLs embedded within emails or websites. 
Accordingly, detecting such URLs through lexical and structural analysis has become a focal point of phishing research.

\subsubsection{Lexical and Host-Based URL Feature Analysis}

One of the pioneering works in this field was presented by Ma et al. \cite{ma2011}, who developed a machine learning framework that used only lexical and host-based features derived directly from the URL string. 

Key lexical features extracted from the URL included length, number of dots, special character counts, and presence of suspicious keywords such as "secure" or "login." Host-based features involved WHOIS data, domain age, and IP address types. These features were input into classifiers like Naïve Bayes, Logistic Regression, and Decision Trees. Despite the simplicity of the features, the models demonstrated strong classification accuracy and adaptability, 
The researchers demonstrated that lexical-only approaches can be highly effective when crafted with well-targeted features, especially in resource-constrained environments.

\begin{figure*}[t]
    \centering
    \includegraphics[width=0.8\textwidth]{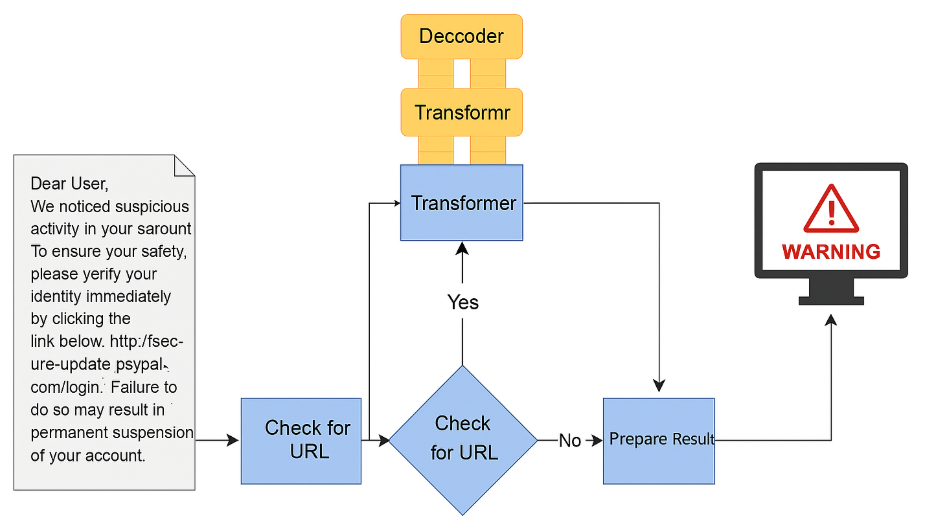}
    \caption{High‐level overview of our dual‐path phishing detection pipeline. The raw email is split into (1) a text stream, tokenized and processed by an encoder–decoder Transformer (inset shows attention weights on the misspelling `sauront'), and (2) a URL stream, subjected to character‐level spoofing analysis (inset highlights `psypal' vs. `paypal'). The resulting embeddings are fused in a decision layer to output a `Phishing' or `Legitimate' label.}

    \label{fig:pipeline}
\end{figure*}

\subsubsection{Deep Learning Approaches for URL Representation}

Recognizing the limitations of manually engineered features, Le et al. \cite{le2018} proposed a deep learning-based solution that automatically learns representations from raw URL strings. Their model leveraged a Character-level Convolutional Neural Network (CNN), treating URLs as sequences of individual characters instead of words. This allowed the system to detect subtle patterns and anomalies such as random character generation, visual spoofing (e.g., "g00gle.com"), or unusual TLDs.

By bypassing the need for feature engineering, the CNN was able to generalize better across different phishing strategies and adapt to emerging threats. 
Experimental results showed that the character-level CNN achieved superior detection accuracy, particularly in identifying zero-day URLs that had not been previously encountered in training data.

This approach exemplifies the power of end-to-end deep learning in cybersecurity, offering an automated pipeline for real-time malicious URL detection. 

\subsection{Integrated Approaches in Phishing Detection}

Isolated feature-based systems—whether focusing on email content or URLs—often fall short when confronting sophisticated phishing attacks that combine multiple forms of deception. Integrated approaches have emerged as a promising strategy to build robust phishing detection systems by fusing linguistic analysis, sender metadata, URL inspection, and even visual components.

\subsubsection{Hybrid Feature Fusion Techniques}

Abdelhamid et al. \cite{abdelhamid2014} introduced a multi-layered detection system that merged sender-based heuristics, lexical content features, and URL inspection into a single framework. Their model extracted linguistic cues such as urgency, misspellings, and suspicious phrases from email text while simultaneously analyzing embedded URLs for characteristics like IP address usage, abnormal domain length, and suspicious TLDs.

In terms of modeling, they utilized an ensemble of Decision Trees, SVM, and Naïve Bayes classifiers, with predictions combined via a majority voting scheme. This ensemble architecture achieved a higher F1-score than any of the individual classifiers alone. Moreover, the system proved resilient to advanced phishing emails that succeeded in bypassing single-layer detection systems, thanks to the diverse range of features under consideration.

\subsubsection{Visual, Structural, and Meta-Information Fusion}

Expanding on the concept of hybridization, Basnet et al. \cite{basnet2008} presented a layered phishing detection model that mimicked how a human might evaluate an email. Their approach divided email analysis into four categories: visual features, link-based characteristics, metadata from headers, and language-based content analysis.

A unique contribution of this work was the inclusion of visual similarity checks—analyzing embedded logos, colors, and layouts to detect phishing emails that mimic branding elements of trusted organizations. 

These integrated approaches highlight the importance of cross-domain feature engineering and ensemble learning in combatting modern phishing threats. By combining content-level, URL-level, and even perceptual indicators, these systems offer comprehensive protection against the evolving tactics employed by attackers.


\section{Methodology}
The objective of our work is to build a robust two-phase system for phishing email detection. Phase 1 utilizes Natural Language Processing (NLP) techniques to analyze the textual content of emails, while Phase 2 applies classical machine learning methods to analyze URLs embedded within emails. The goal is to leverage both semantic and structural phishing cues to enhance overall detection accuracy.

\subsection{Data Collection}
To effectively train and evaluate our system, two distinct datasets were collected. For Phase 1, a diverse set of phishing and legitimate emails was gathered from well-known repositories, including Nazario, Enron, SpamAssassin, and CEAS 08. Initially, the merged dataset contained over 82,000 entries. For Phase 2, a large-scale URL dataset was sourced from Kaggle \cite{kaggleurls}, comprising over 641,000 labeled URLs. The labels encompassed several classes such as phishing, malware, defacement, and benign.

\subsection{Data Preprocessing}
\subsubsection{Email Text Preprocessing (Phase 1)}
The collected email dataset underwent comprehensive preprocessing steps to ensure quality and consistency. Initially, null and duplicate entries were removed, and excessively short messages were eliminated. Subsequently, all text entries were standardized by converting them to lowercase. To expedite experimentation without sacrificing representativeness, the dataset was reduced from its original size of 82,000 to a balanced subset of 7,500 entries, each clearly labeled as phishing (1) or benign (0).

\subsubsection{URL Preprocessing (Phase 2)}
The URL dataset was similarly subjected to rigorous preprocessing. Non-benign labels (phishing, malware, defacement) were aggregated and assigned a single malicious class (1), with benign URLs labeled as 0. Malformed URLs and those shorter than 10 characters were discarded, alongside null and duplicate entries. URLs were converted to lowercase to maintain uniformity. Finally, the dataset was subsampled to 20,000 entries, ensuring computational efficiency while maintaining representational diversity.

\subsection{Feature Engineering}
Effective representation of the data was crucial for optimal model performance. Different feature engineering techniques were employed for the textual and URL analyses to best capture their unique characteristics.

\subsubsection{Tokenization for NLP (Phase 1)}
Textual data for the NLP-based classification were tokenized using Hugging Face's AutoTokenizer. Each transformer model employs its specific tokenizer (e.g., WordPiece tokenizer for BERT), standardizing inputs into consistent subword tokens. Special tokens ([CLS], [SEP]) were added, and sequences were padded or truncated to a fixed maximum length of 256 tokens. For instance, the sentence \texttt{"Please verify your account to avoid suspension."} is tokenized as \texttt{[’[CLS]’, ’please’, ’verify’, ’your’, ’account’, ’to’, ’avoid’, ’suspension’, ’.’, ’[SEP]’]}. Resultant token sequences were converted into PyTorch-compatible tensors (input ids, attention masks, and labels), thus ensuring seamless integration with the models.

\subsubsection{Vectorization for URL Analysis (Phase 2)}
URLs were numerically encoded using the \texttt{TfidfVectorizer} from scikit-learn, configured for character-level analysis. The analyzer was set to generate character-level n-grams ranging from 2 to 6 characters. This approach effectively captured significant lexical patterns indicative of phishing URLs, such as common deceptive terms (e.g., "login," "secure," "verify") and suspicious domain indicators (e.g., ".ru," ".xyz," or IP address-based URLs). The output was a sparse TF-IDF feature matrix suitable for classical machine learning methods.

\subsection{Model Design}
The proposed system comprises two complementary modules: an NLP-based email classification component and a URL-based link analysis component. Each component functions independently, but they can also be integrated into a multisignal ensemble in future iterations. This modular design aligns closely with recent research trends emphasizing interpretability in AI-driven phishing detection \cite{alsubaiey2024phishing}.

\subsubsection{NLP-Based Email Classification}
We selected five transformer-based architectures from Hugging Face’s library due to their proven performance in textual classification tasks. The chosen models were: \texttt{bert-base-uncased}, \texttt{distilbert-base-uncased}, \texttt{microsoft/MiniLM-L12-H384-uncased}, \texttt{huawei-noah/TinyBERT\_General\_4L\_312D}, and \texttt{prajjwal1/bert-tiny}. Each model was fine-tuned on the 7,500-entry preprocessed email dataset.

\subsubsection{URL-Based Link Analysis}
Two classical machine learning algorithms were implemented for URL classification based on the TF-IDF vectorized data: Logistic Regression and Random Forest. Random Forest was set with 100 estimators, while Logistic Regression utilized default parameters to ensure robust baseline comparisons.

\subsection{Training Procedure}
For the NLP component, each transformer-based model was trained for 3 epochs using Hugging Face's Trainer API. A learning rate of 2e-5 and batch size of 8 were employed during training, with checkpoint selection based on validation loss to prevent overfitting.

For the URL component, an 80/20 train-test split of the vectorized URL dataset was used to train and evaluate both the Logistic Regression and Random Forest classifiers. Training followed standard procedures without further adjustments, ensuring replicability and fairness in comparisons.

\subsection{Hyperparameter Tuning}
Transformer models utilized hyperparameters based on common best practices (learning rate: 2e-5, batch size: 8), informed by preliminary experiments. Classical machine learning models employed default hyperparameters, except for Random Forest’s number of estimators, deliberately set to 100 based on preliminary evaluations to balance performance and computational efficiency.

\subsection{Evaluation}
The trained models from both phases were rigorously evaluated to ensure accuracy, reliability, and generalizability. The transformer-based classifiers were assessed primarily on classification accuracy and F1-score. Classical URL classifiers were evaluated similarly, with additional consideration of computational complexity and feature interpretability. The evaluation procedures included thorough comparative analyses across models to determine optimal selections for practical deployment scenarios.

\begin{table}[t]
  \centering
  \caption{Classification performance of transformer-based NLP models on the phishing email dataset, evaluated by accuracy, F1 score, precision, and recall.}
  \label{tab:nlp_model_comparison}
  \begin{tabular}{lcccc}
    \toprule
    Model            & Accuracy & F1 score & Precision & Recall \\
    \midrule
    BERT-Base        & 0.994    & 0.995    & 0.993     & 0.996  \\
    DistilBERT       & 0.989    & 0.989    & 0.988     & 0.989  \\
    MiniLM-L12       & 0.986    & 0.987    & 0.991     & 0.982  \\
    Noah Tiny BERT   & 0.964    & 0.966    & 0.953     & 0.980  \\
    BERT Tiny        & 0.948    & 0.950    & 0.943     & 0.955  \\
    \bottomrule
  \end{tabular}
\end{table}

\begin{table}[h]
  \centering
  \caption{Performance of URL‐based classifiers on our test set (TF-IDF + Random Forest)}
  \label{tab:url_classifiers}
  \begin{tabular}{lcc}
    \toprule
    \textbf{Model} & \textbf{Accuracy} & \textbf{F1 Score} \\
    \midrule
    TF-IDF + Logistic Regression       & 0.9255            & 0.926             \\
    TF-IDF + Random Forest             & \textbf{0.94675}  & \textbf{0.947}    \\
    \bottomrule
  \end{tabular}
\end{table}

\section{Results}
\subsection{Overall Performance}
The effectiveness of the proposed phishing detection framework was evaluated separately for the NLP-based email classification (Phase 1) and the URL-based link analysis (Phase 2). Table~\ref{tab:nlp_model_comparison} and Table~\ref{tab:url_classifiers} summarize the quantitative performance metrics of these two phases, respectively. Key insights and comparative analyses follow in detail.

\subsection{NLP Classification Results}
Five transformer-based NLP architectures were fine-tuned on the curated email dataset of 7,500 entries, and their performances are compared in Table~\ref{tab:nlp_model_comparison}. BERT-Base demonstrated superior performance, achieving the highest accuracy (0.994) and F1 score (0.995). However, its training was computationally demanding, resulting in the longest training duration among the tested models.

DistilBERT, in contrast, delivered nearly comparable accuracy (0.989) and F1 score (0.989) but required significantly less computational resources and training time. Thus, DistilBERT offers an optimal balance between computational efficiency and detection performance, making it particularly suitable for scenarios where computational resources are limited.

Smaller-scale models, including MiniLM-L12, Noah Tiny BERT, and BERT Tiny, exhibited comparatively lower performance scores, with accuracy ranging from 0.948 (BERT Tiny) to 0.986 (MiniLM-L12). Despite their reduced accuracy, these compact models trained exceptionally quickly, underscoring their potential suitability for deployment in edge computing environments or resource-constrained applications.

These results collectively demonstrate that transformer-based models, particularly BERT-Base and DistilBERT, provide robust solutions for the semantic detection of phishing content in emails, with DistilBERT recommended as the best compromise between performance and efficiency.

\subsection{URL-Based Detection Results}
Table~\ref{tab:url_classifiers} presents the performance of classical machine learning models trained on the TF-IDF vectorized URL dataset. The Random Forest classifier achieved the highest accuracy (0.94675) and F1 score (0.947), notably surpassing Logistic Regression (accuracy: 0.9255, F1 score: 0.926). This performance gap suggests that Random Forest more effectively captures intricate structural patterns inherent in malicious URLs. Furthermore, both models operated without explicit manual feature extraction beyond TF-IDF vectorization, emphasizing their computational efficiency and scalability.

\subsection{Combined Insights}
While the NLP-based and URL-based classifiers were evaluated independently, their complementary strengths provide a strong rationale for future integration. The NLP classifiers demonstrated effectiveness in capturing semantic indicators of social engineering, whereas the URL classifiers excelled at identifying structural features characteristic of malicious links. A combined ensemble approach, potentially utilizing weighted decision fusion, could leverage these complementary strengths to achieve higher overall detection rates and reduce false-negative rates.

\subsection{Limitations and Future Work}
Several limitations of the current study warrant consideration in future research. The NLP-based component was evaluated on a limited dataset of 7,500 emails, suggesting potential improvements in generalizability if trained on larger and more diverse corpora. Additionally, testing more computationally intensive transformer models such as RoBERTa or DeBERTa could lead to further performance enhancements.

Integration of the two independently evaluated phases into a unified hybrid or ensemble system presents another avenue for enhancing performance, particularly by addressing weaknesses inherent to each approach individually. Finally, evaluation of the system against live phishing campaigns or real-world phishing datasets would significantly enhance confidence in its practical applicability and robustness.

Addressing these areas in future iterations will facilitate the development of a more accurate, scalable, and robust phishing detection system suitable for real-world email security deployments.

\section{Conclusion}
This paper presented a two-phase system for phishing email detection that combines Natural Language Processing (NLP)-based text classification with URL-based link analysis. In Phase 1, transformer-based models such as BERT and DistilBERT were fine-tuned to classify email content with high accuracy, leveraging semantic understanding to detect social engineering attempts. In Phase 2, a TF-IDF vectorization approach was used to extract character-level features from embedded URLs, which were then classified using classical machine learning models. The Random Forest classifier notably outperformed Logistic Regression, demonstrating strong capability in capturing structural patterns within URLs.

The results showed that DistilBERT offered the best trade-off between performance and efficiency for NLP tasks, while the URL-based Random Forest model provided high accuracy with minimal preprocessing. Together, these findings support the feasibility of a hybrid detection architecture that leverages the strengths of both content and link analysis. Additionally, expanding the datasets and including multilingual capabilities may further enhance the system's robustness and generalizability in diverse deployment environments.
% ...

\bibliographystyle{IEEEtran}
\bibliography{bib}

\begin{thebibliography}{10}
\providecommand{\url}[1]{#1}
\csname url@samestyle\endcsname
\providecommand{\newblock}{\relax}
\providecommand{\bibinfo}[2]{#2}
\providecommand{\BIBentrySTDinterwordspacing}{\spaceskip=0pt\relax}
\providecommand{\BIBentryALTinterwordstretchfactor}{4}
\providecommand{\BIBentryALTinterwordspacing}{\spaceskip=\fontdimen2\font plus
\BIBentryALTinterwordstretchfactor\fontdimen3\font minus \fontdimen4\font\relax}
\providecommand{\BIBforeignlanguage}[2]{{%
\expandafter\ifx\csname l@#1\endcsname\relax
\typeout{** WARNING: IEEEtran.bst: No hyphenation pattern has been}%
\typeout{** loaded for the language `#1'. Using the pattern for}%
\typeout{** the default language instead.}%
\else
\language=\csname l@#1\endcsname
\fi
#2}}
\providecommand{\BIBdecl}{\relax}
\BIBdecl

\bibitem{aag_it_phishing_stats}
{AAG IT}, ``The latest phishing statistics (updated january 2025),'' Online, Jan. 2025, [Online]. Available: \url{https://aag-it.com/the-latest-phishing-statistics/}.

\bibitem{kyaw2024}
P.~H. Kyaw, J.~Gutierrez, and A.~Ghobakhlou, ``A systematic review of deep learning techniques for phishing email detection,'' \emph{Electronics}, vol.~13, no.~19, p. 3823, 2024.

\bibitem{thakur2023}
K.~Thakur, M.~L. Ali, M.~A. Obaidat, and A.~Kamruzzaman, ``A systematic review on deep-learning-based phishing email detection,'' \emph{Electronics}, vol.~12, no.~21, p. 4545, 2023.

\bibitem{atlam2023}
H.~F. Atlam and O.~Oluwatimilehin, ``Business email compromise phishing detection based on machine learning: A systematic literature review,'' \emph{Electronics}, vol.~12, no.~1, p.~42, 2023.

\bibitem{rayala2023}
R.~Rayala, R.~Kuppa, S.~Pasumarthi, and S.~R. Karthik, ``Malicious url detection using logistic regression,'' TechRxiv, 2023, [Online]. Available: \url{https://doi.org/10.36227/techrxiv.14725539.v1}.

\bibitem{sahami1998}
M.~Sahami, S.~Dumais, D.~Heckerman, and E.~Horvitz, ``A bayesian approach to filtering junk e-mail,'' in \emph{Learning for Text Categorization: Papers from the 1998 Workshop}.\hskip 1em plus 0.5em minus 0.4em\relax Madison, Wisconsin, USA: AAAI Press, 1998, pp. 55--62, aAAI Technical Report WS-98-05.

\bibitem{drucker1999}
H.~Drucker, D.~Wu, and V.~N. Vapnik, ``Support vector machines for spam categorization,'' \emph{IEEE Transactions on Neural Networks}, vol.~10, no.~5, pp. 1048--1054, 1999.

\bibitem{zhang2020}
Y.~Zhang, J.~Peng, and C.~Zhang, ``Email phishing detection using bert-based deep learning,'' in \emph{Proc. IEEE Int. Conf. Computer Communications and Networks (ICCCN)}, Honolulu, HI, USA, Aug. 2020, pp. 1--7.

\bibitem{abdulnabi2021}
I.~AbdulNabi and Q.~Yaseen, ``Spam email detection using deep learning techniques,'' \emph{Procedia Comput. Sci.}, vol. 184, pp. 853--858, 2021.

\bibitem{sahingoz2019}
O.~K. Sahingoz, E.~Buber, O.~Demir, and B.~Diri, ``Machine learning based phishing detection from urls,'' \emph{Expert Syst. Appl.}, vol. 117, pp. 345--357, Mar. 2019.

\bibitem{chandrasekaran2006}
R.~Chandrasekaran, K.~Narayanan, and S.~Upadhyaya, ``Phishing email detection based on structural properties,'' in \emph{Proc. NYS Cyber Security Conf.}, Albany, NY, USA, 2006, pp. 1--7.

\bibitem{ma2011}
J.~Ma, L.~K. Saul, S.~Savage, and G.~M. Voelker, ``Learning to detect malicious urls,'' \emph{ACM Trans. Intell. Syst. Technol.}, vol.~2, no.~3, pp. 1--24, May 2011.

\bibitem{le2018}
T.~Le, A.~Markham, and Y.~Zhang, ``Urlnet: Learning a url representation with deep learning for malicious url detection,'' in \emph{Proc. IEEE Conf. Data Mining Workshops (ICDMW)}, Singapore, Nov. 2018, pp. 275--282.

\bibitem{abdelhamid2014}
N.~Abdelhamid, A.~Ayesh, and F.~Thabtah, ``Phishing detection based on hybrid features,'' in \emph{Proc. IEEE Int. Conf. Cybercrime and Computer Forensic (ICCCF)}, Vancouver, Canada, Jun. 2014, pp. 1--6.

\bibitem{basnet2008}
R.~Basnet, A.~H. Sung, and Q.~Liu, ``Rule-based phishing attack detection in email,'' in \emph{Proc. Int. Conf. Security and Management (SAM)}, Las Vegas, NV, USA, Jul. 2008, pp. 1--7.

\bibitem{kaggleurls}
Kaggle, ``Malicious urls dataset,'' Online, Jul.~23 2021, [Online]. Available: \url{https://www.kaggle.com/datasets/sid321axn/maliciousurls-dataset}.

\bibitem{alsubaiey2024phishing}
A.~Al-Subaiey, M.~Al-Thani, N.~A. Alam, K.~F. Antora, A.~Khandakar, and S.~A.~U. Zaman, ``Novel interpretable and robust web-based ai platform for phishing email detection,'' arXiv preprint arXiv:2405.11619, May 2024, [Online]. Available: \url{https://arxiv.org/abs/2405.11619}.

\end{thebibliography}

\end{document}